\documentclass[aps,superscriptaddress,twocolumn,a4paper,10pt]{revtex4-2}
\usepackage[colorlinks=true, pdfstartview=FitV, linkcolor=blue, citecolor=red, urlcolor=black]{hyperref}
\usepackage{graphicx}
\usepackage[all]{xy}
\usepackage{amsmath}
\usepackage{amssymb}
\usepackage{tensor}
\usepackage{orcidlink}
\usepackage{amsthm}
\usepackage{comment}
\newcommand{\be}{\begin{equation}}
\newcommand{\ee}{\end{equation}}
\newcommand{\ben}{\begin{eqnarray}}
\newcommand{\een}{\end{eqnarray}}
\newcommand{\bes}{\begin{subequations}}
\newcommand{\ees}{\end{subequations}}
\def\bal#1\eal{\begin{align}#1\end{align}}

\newcommand{\bfi}{\begin{figure}}
\newcommand{\efi}{\end{figure}}
\newcommand{\bc}{\begin{center}}
\newcommand{\ec}{\end{center}}

\newcommand{\sech}{{\rm sech}}

\newcommand{\LL}{{\cal L}}

\newcommand{\Sc}{{\cal S}}

\newcommand{\qandq}{\quad {\rm and} \quad }
 
\begin{document}

\title{Bound states around vacuum in scalar ModMax model}
\author{F.A. Brito\,\orcidlink{0000-0001-9465-6868}}\email{fabrito@df.ufcg.edu.br}
\affiliation{Departamento de F\'{\i}sica, Universidade Federal de Campina Grande,  58109-970 Campina Grande, PB, Brazil}
\author{M.A. Marques\,\orcidlink{0000-0001-7022-5502}}
        \email[]{marques@cbiotec.ufpb.br}\affiliation{Departamento de Biotecnologia, Universidade Federal da Para\'\i ba, 58051-900 Jo\~ao Pessoa, PB, Brazil}
\author{R. Menezes\,\orcidlink{0000-0002-9586-4308}}
     \email[]{rmenezes@dcx.ufpb.br}\affiliation{Departamento de Ci\^encias Exatas, Universidade Federal
da Para\'{\i}ba, 58297-000 Rio Tinto, PB, Brazil}\affiliation{Departamento de F\'{\i}sica, Universidade Federal de Campina Grande,  58109-970 Campina Grande, PB, Brazil}

\author{E. Passos\,\orcidlink{0000-0003-1718-6385}}\email{passos@df.ufcg.edu.br}
\affiliation{Departamento de F\'{\i}sica, Universidade Federal de Campina Grande,  58109-970 Campina Grande, PB, Brazil}

\begin{abstract}
In this work, we consider a two-dimensional scalar field model inspired by the dimensional reduction of a four-dimensional ModMax theory. Upon projecting out the 4D theory down to a 2D theory we obtain a theory which presents a constant electric field and two scalar fields. In order to investigate kinks, we include the presence of a potential and consider the static case with one of the fields in the vacuum, showing that the solutions for the non-uniform field can be mapped into the ones arising from the canonical model. By studying the linear stability of the model, we show that fluctuations around the uniform field are described by a Sturm-Liouville eigenvalue equation whose weight function depends on the non-uniform solution and the parameter of the ModMax model. Remarkably, the presence of the aforementioned weight may bring bound states to light, contrary to what occurs in the canonical model.

\end{abstract} 

\maketitle

In Ref.~\cite{modmax1}, an extension of Maxwell's equations via a non-linear model invariant under both duality rotations and conformal transformations, called ModMax, was proposed. Its action has the form $\Sc = \!\int\! d^4x\,\LL$, where the Lagrangian density is
\be\label{LLmodmax}
\LL=\cosh \gamma\, S+ \sinh \gamma \,\sqrt{S^2+P^2},
\ee
in which $S$ and $P$ represent the Maxwell invariants,
\be
S=-\frac14 F_{\Tilde{\mu}\Tilde{\nu}}F^{\Tilde{\mu}\Tilde{\nu}}\qandq P=-\frac14 F_{\Tilde{\mu}\Tilde{\nu}}\,{^*}F^{\Tilde{\mu}\Tilde{\nu}}.
\ee
In the above expressions, $F_{\Tilde{\mu}\Tilde{\nu}} = \partial_{\Tilde{\mu}} A_{\Tilde{\nu}} - \partial_{\Tilde{\nu}} A_{\Tilde{\mu}}$ and its dual is represented by ${^*}F_{\Tilde{\mu}\Tilde{\nu}} = \epsilon_{\Tilde{\mu}\Tilde{\nu}\Tilde{\lambda}\Tilde{\rho}}F^{\Tilde{\lambda}\Tilde{\rho}}$, with $\epsilon_{\Tilde{\mu}\Tilde{\nu}\Tilde{\lambda}\Tilde{\rho}}$ representing the Levi-Civita symbol, such that $\epsilon_{0123}=-1$. This model is defined in $3+1$ dimensions, so the Greek indices with tilde runs from $0$ to $3$ and the metric tensor is $\eta_{\Tilde{\mu}\Tilde{\nu}}=\text{diag}(+,-,-,-)$. In the Lagrangian density \eqref{LLmodmax}, $\gamma$ is a real parameter which recovers the standard Maxwell electrodynamics for $\gamma=0$. This model has been widely investigated recently; see, for instance, Refs.~\cite{modmax1,modmax2,modmax3,modmax4,modmax5,modmax6,modmax7,modmax8,modmax9,modmax10,modmax11,modmax12,modmax13}.

Since we are interested in investigating the presence of kinks, we get inspiration from Refs.~\cite{reddim1,reddim2,reddim3,reddim4} and make the dimensional reduction from $4$D to $2$D. By following the prescription in the aforementioned references, we consider that $\partial_2=\partial_3=0$ when acting in the fields, and $A^{\Tilde{\mu}} = (A^\mu, \chi, \phi)$, $\chi \equiv A_2$ and $\phi \equiv A_3$. The non-tilde Greek indices span from $0$ to $1$. This procedure leads us to 
\be
S= -\frac14 F_{\mu\nu}F^{\mu\nu} + \frac12  \partial_\mu\phi \partial^\mu \phi + \frac12  \partial_\mu\chi \partial^\mu \chi
\ee
and
\be
P=\varepsilon_{\mu\nu}\partial^\mu \phi \partial^\nu \chi,
\ee
in which we have used $\epsilon_{\Tilde{\mu}\Tilde{\nu}23} = \epsilon_{\mu\nu}$, such that $\epsilon_{01}=-1$. The 2D Lagrangian density is
\be\label{LLproj}
\begin{aligned}
   \LL=\cosh \gamma\, \left(\frac12E^2+X\right)+ \sinh \gamma\, \sqrt{\left(\frac12E^2+X\right)^2+Y^2}.
\end{aligned}
\ee
The dynamics in the above expression is driven by the terms $X$ and $Y$, which we define as 
\be\label{XY}
\begin{aligned}
  &  X=\frac12 \partial_\mu \phi \partial^\mu \phi+\frac12 \partial_\mu \chi \partial^\mu \chi =\frac12{\dot \phi}^2-\frac12{\phi^{\prime}}^2 +
\frac12{\dot \chi}^2-\frac12{\chi^{\prime}}^2 \\
   & Y=\varepsilon_{\mu\nu}\partial^\mu \phi \partial^\nu \chi = \phi^\prime \dot \chi -\chi^\prime \dot \phi,
\end{aligned}
\ee
in which the dots and primes stand for the derivatives with respect to time and space, respectively. For time-dependent fields, the change $x\to-x$ and/or $t\to-t$ keeps $X$ unchanged. However, $Y$ has a distinct behavior: in order to keep it invariant, one must perform \emph{both} transformations simultaneously, as there is a change of sign if just one of the operations are done. It is also worth commenting that scalar fields depending exclusively on the space \emph{or} the time leads to vanishing $Y$.

As it is known, the ModMax model has subtleties in points where the derivatives of the fields vanish. So, one must be careful when dealing with it. To avoid issues in the aforementioned points, we consider that the system has a constant electric field $E$ and define $C\equiv-F_{\mu\nu}F^{\mu\nu}/4=E^2/2$, obeying $P>0$. Since we intend to investigate the presence of kinks in scalar field models, we must include a potential. With this motivation, we introduce the Lagrangian density
\be\label{LLU}
\begin{aligned}
   \LL=\cosh \gamma\, X+ \sinh \gamma\, \sqrt{\left(C+X\right)^2+Y^2} - U(\phi,\chi),
\end{aligned}
\ee
where $U(\phi,\chi)$ represents the potential. By comparing the Lagrangian density \eqref{LLU} with \eqref{LLproj}, we see that a constant term was absorbed in the potential, for simplicity. We remark that the above Lagrangian density \emph{is not} obtained directly from the dimensional reduction of \eqref{LLmodmax}, as it includes a potential. The above model takes into account the effects of a classical $2D$ root-TT deformation, previously introduced in \cite{ferkoprl}, which introduces the non-analytic kinetic mixing in scalar field models without modifying the potential.

We are interested in studying the specific situation in which the fields are decoupled in the potential and one of the fields is in the vacuum, so we take $U(\phi,\chi)=V(\phi)+m^2\chi^2/2$, with $m$ representing a mass, and work with the Lagrangian density
\be\label{LL}
\begin{aligned}
   \LL=\cosh \gamma X+ \sinh \gamma \sqrt{\left(C+X\right)^2+Y^2} - V(\phi)-\frac{m^2}{2}\chi^2,
\end{aligned}
\ee
where $V(\phi)$ is the potential which controls the self-interaction of the field responsible for giving rise to defect structures. From now on, we call \eqref{LL} scalar ModMax model.

By varying the action with respect to the
scalar fields, we get the following equations of
motion
\bes\label{eom}
\bal\label{eomphi}
\partial^\mu \Pi^\phi_\mu +V_\phi = 0,\\ \label{eomchi}
\partial^\mu \Pi^\chi_\mu +m^2\chi = 0
\eal
\ees
where $V_\phi=dV/d\phi$. In the above expression, we have defined the generalized momenta associated to the scalar fields as
\bes\bal
\Pi^\phi_\mu = \LL_X\partial_\mu \phi + \LL_Y\varepsilon_{\mu\nu} \partial^\nu \chi,\\ \Pi^\chi_\mu =\LL_X \partial_\mu \chi - \LL_Y\varepsilon_{\mu\nu} \partial^\nu \phi.
\eal
\ees
The factors $\LL_X=\partial \LL/\partial X$ and $\LL_Y=\partial \LL/\partial Y$ in the above expressions are given by
\bes\label{LXLY}
\bal
&\LL_X = \cosh \gamma + \sinh \gamma \,\frac{C+X}{\sqrt{\left(C+X\right)^2+Y^2}},\\&\LL_Y=\sinh \gamma \,\frac{Y}{\sqrt{\left(C+X\right)^2+Y^2}}.
\eal
\ees
We see that the equation of motion \eqref{eomchi} supports the uniform solution $\chi(x)=0$, which we take from now on. To investigate the presence of kinks in the model \eqref{LL}, we consider static configurations, i.e., $\phi=\phi(x)$. In this situation, the kinetic terms \eqref{XY} become $X=-{\phi^{\prime}}^2/2$ and $Y=0$. The equation of motion \eqref{eomphi} reads  $e^{\gamma}\phi^{\prime\prime}  =V_\phi$. We may define the coordinate $y=e^{-\gamma/2}x$ to write it in the form
\be\label{eomstatic}
\phi_{yy}=V_\phi,
\ee
where $\phi_{yy}=d^2\phi/dy^2$. The presence of the coordinate $y$ brings an interesting perspective to light, related to the presence of a modified metric that maps the solutions of the equations of motion associated to TT- or root TT-deformed models; see Refs.~\cite{modmetric1,modmetric2}. This equation has the same form of its corresponding one in the canonical model, $\LL=\partial_\mu\phi\partial^\mu\phi/2-V(\phi)$, for static $\phi$. Therefore, we can map the known solutions of the canonical model into our current one. Since we are interested in kink structures, we must solve the above equation considering that the solution connects two neighbor minima of the potential \cite{vachaspati}. In this direction, we take the sine-Gordon model $V(\phi)=(1/2)\cos^2\phi$. This potential has a family of minima. Its central sector, defined by the interval $[-\pi/2,\pi/2]$, supports the kink solution
\be\label{sol}
\phi(y) =\arcsin(\tanh(y))
\ee 
in its respective equation of motion \eqref{eomstatic}.

Considering $\chi(x)=0$ and a solution of the equation of motion \eqref{eomstatic}, $\phi(x)$, let us investigate their behavior in the presence of small fluctuations, in the lines of Refs.~\cite{trilogia1,trilogia2,stabkink}. We then write the perturbed fields as $\phi(x,t) = \phi(x) + \sum_k\eta^{(k)}(x)\cos\big(\omega^{(k)}_\phi t\big)$ and $\chi(x,t) =\sum_k\xi^{(k)}(x)\cos\big(\omega^{(k)} t\big)$. By using the equations of motion \eqref{eom}, we get that the fluctuations around $\phi$ are governed by the eigenvalue equation $-\eta_{yy} + V_{\phi\phi}(y)\eta= \omega_\phi^2e^{\gamma}\eta$. Notice that $\eta$ does not depend on the fluctuations of the $\chi$ field; this occurs because we are considering that it is in the vacuum ($\chi=0$). On the other hand, the fluctuations around the vacuum of $\chi$ depend on the other field; they are governed by the Sturm-Liouville equation  
\be\label{stabchi}
-\xi_{yy} + m^2\xi(y)= \omega^2\sigma(y)\xi(y)
\ee
where we have omitted the index $k$ for notational simplicity and the function $\sigma(y)$ is a weight, given by
\be\label{sigma}
\sigma(y)=\cfrac{2\sinh\gamma\,\phi_y^2}{2Ce^{\gamma}-\phi_y^2}+ e^{\gamma}.
\ee
For $\gamma=0$, the above expression simplifies to $\sigma(y)=1$, so the stability equation \eqref{stabchi} reduces to the homogeneous Helmholtz equation, $-\xi_{yy}= (\omega^2-m^2)\xi$. In this specific situation, one can only find solutions for $\omega^2>m^2$ and there are no bound states; this is a well-known result for the canonical model. Therefore, modifications of the eigenvalue spectrum associated to vacuum solutions are solely introduced by the ModMax model via the parameter $\gamma$. The above weight function enters the normalization condition for the modes, which is given by $\int_{-\infty}^{\infty}dy\,\sigma(y)\xi^2(y)=1$.

The presence of the weight function induced by the ModMax model in \eqref{stabchi} leads to an interesting situation in the study of the linear stability. Since $\sigma(y)$ depends on $\phi$, it behaves non-uniformly over the space. This feature allows us to find bound and resonant states, contrary to what occurs in the canonical model.

For a general $\gamma$, any solution $\phi(y)$ connecting two neighbor minima of the potential supports vanishing derivative at infinity. Therefore, for $y\to\pm\infty$, we have $\sigma(y)\to \sigma_\infty\equiv e^\gamma$, which is the minimum (maximum) value of $\sigma$ for positive (negative) $\gamma$. This makes the fluctuations $\xi$ have an asymptotic behavior very similar to the one associated with the aforementioned homogeneous Helmholtz equation. On one hand, if the fluctuations obey the inequality $\omega^2>m^2/\sigma_{\infty}$, then they are not bound states. Remarkably, on the other hand, if $\omega^2<m^2/\sigma_{\infty}$, bound states may exist for positive $\gamma$. This occurs because there are points in the space, which we call $y_*$, that are compatible with $\omega^2> m^2/\sigma(y_*)$.

\begin{figure}[t!]
    \centering
\includegraphics[width=0.5\linewidth]{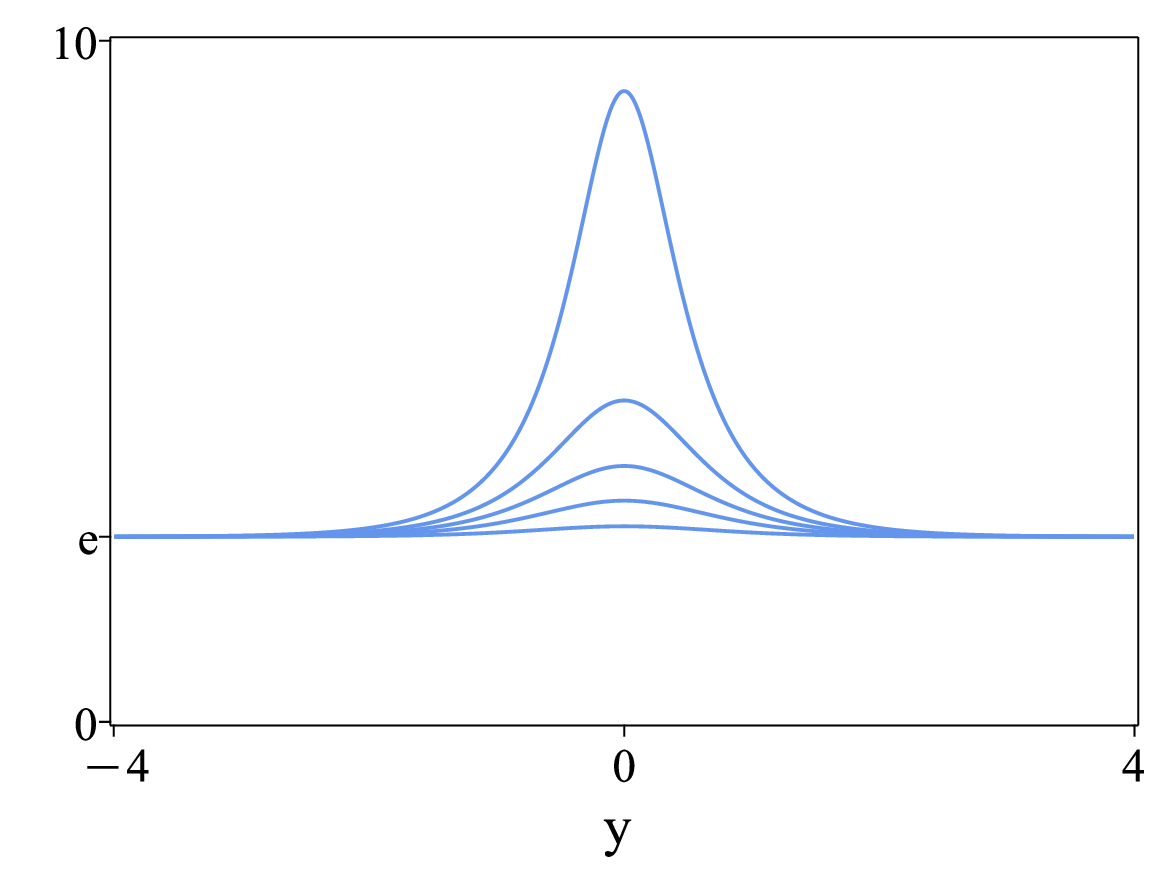}\includegraphics[width=0.5\linewidth]{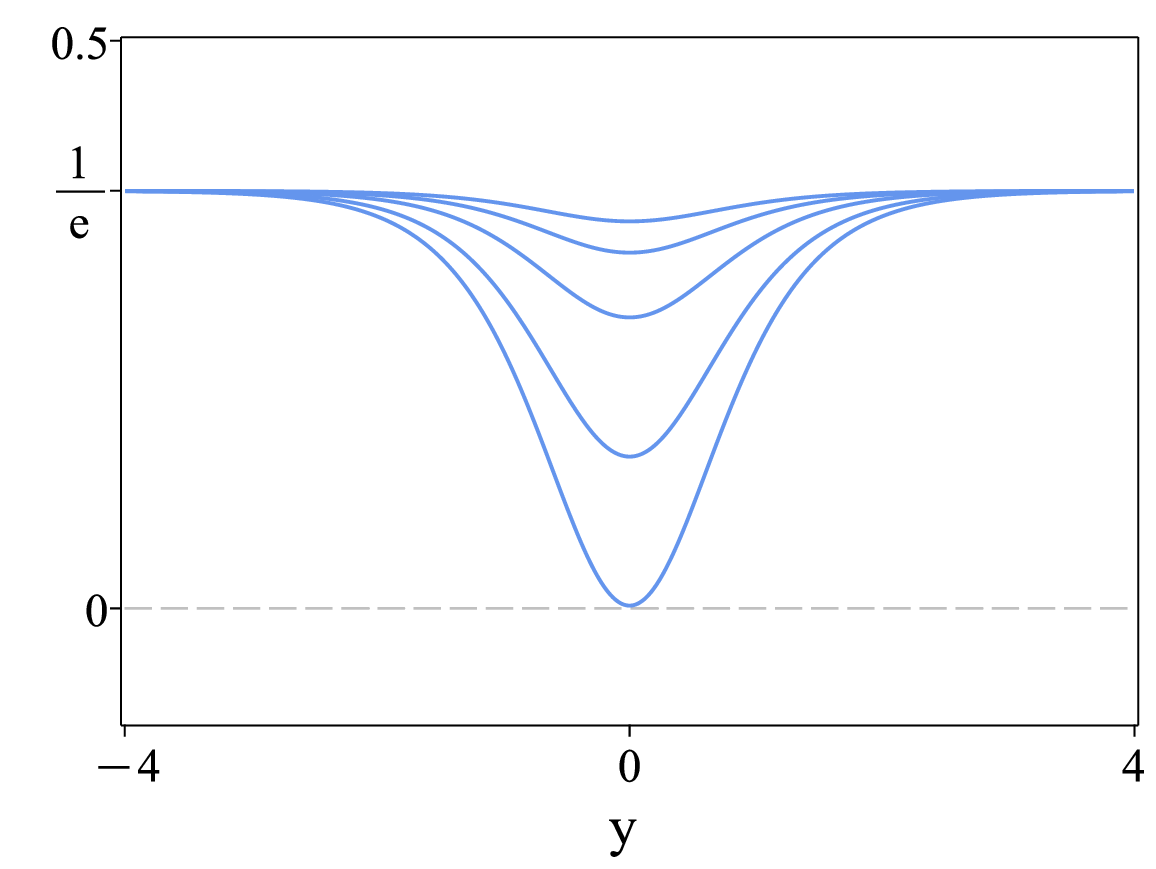}
    \caption{The weight function \eqref{sigmasol} for some values of $\gamma$ and $C$. In the left panel, we depict it for $\gamma=1$ and $C=0.25,0.4,0.6,1$ and $3$. In the right panel, we display it for $\gamma=-1$ and $C=10.01, 15,30,60,120$. The maximum (minimum) goes up (down) as $C$ gets smaller in the left (right) panel.}
    \label{figsigma1}
\end{figure}

For the solution in Eq.~\eqref{sol}, the weight function \eqref{sigma} takes the form
\be\label{sigmasol}
\sigma(y)=\cfrac{2\sinh\gamma\,\sech^2(y)}{2Ce^{\gamma}-\sech^2(y)}+ e^{\gamma}.
\ee
To ensure that $\sigma(y)$ is non negative, we take $C>e^{-3\gamma}/2$. Also, to avoid divergence, we must impose $C>e^{-\gamma}/2$. Therefore, we must take $C>e^{-\gamma}/2$ for $\gamma>0$ and $C>e^{-3\gamma}/2$ for $\gamma<0$ to have an appropriate  $\sigma(y)$. We display the above function in Fig.~\ref{figsigma1}. It has a bell shape, with the extreme located at $y=0$, where $\sigma(0)=2\sinh\gamma/(2Ce^\gamma-1) + e^\gamma$. This point is a maximum for $\gamma>0$ and a minimum for $\gamma<0$. Asymptotically, for $x\to\infty$, we have $\sigma(y)\approx e^\gamma +  4K^{-1}\sinh\gamma\,e^{-\gamma-2y} + {\cal O}[e^{-4y}]$.

\begin{figure}[t!]
    \centering
    \includegraphics[width=0.33\linewidth]{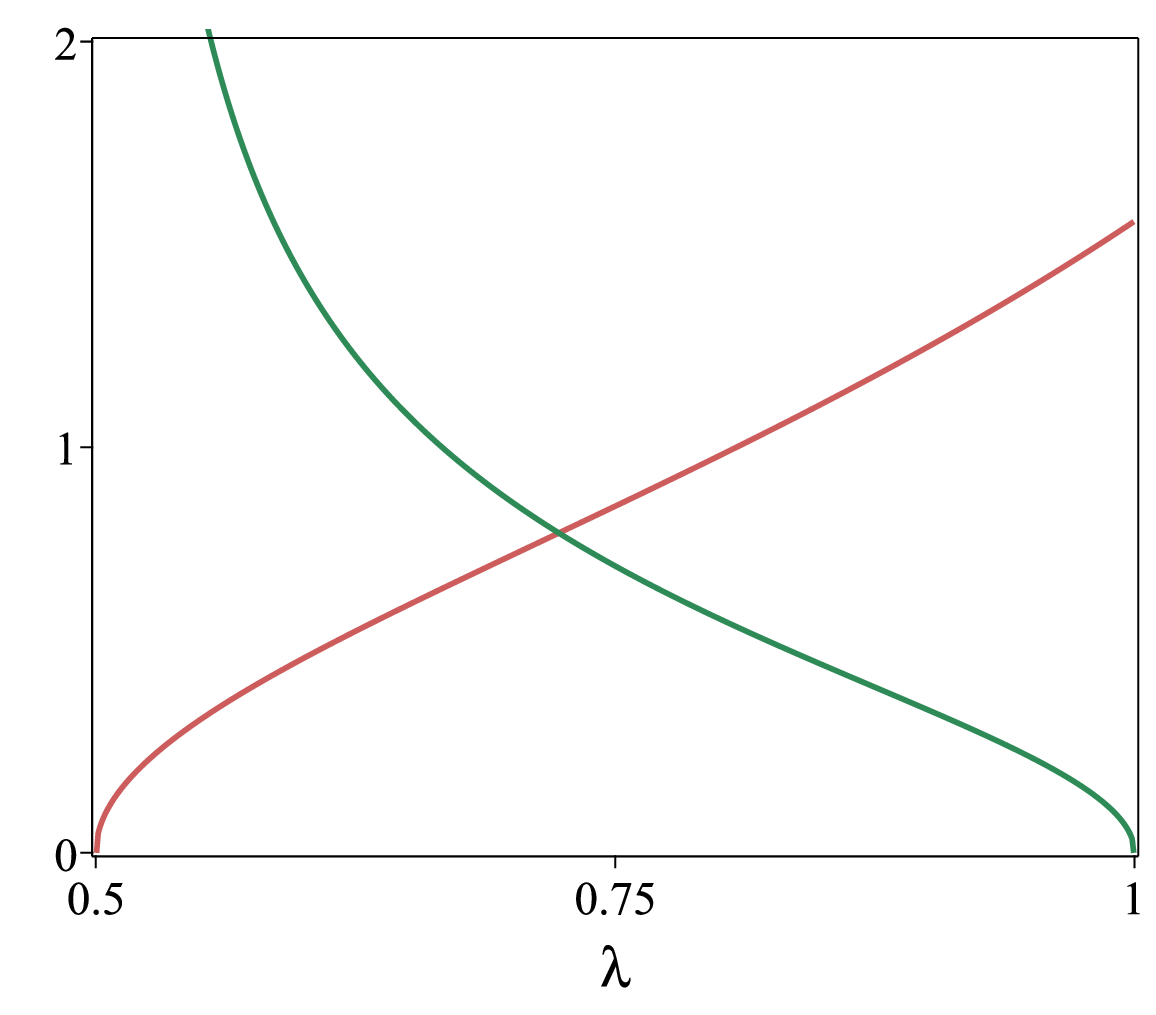}\includegraphics[width=0.33\linewidth]{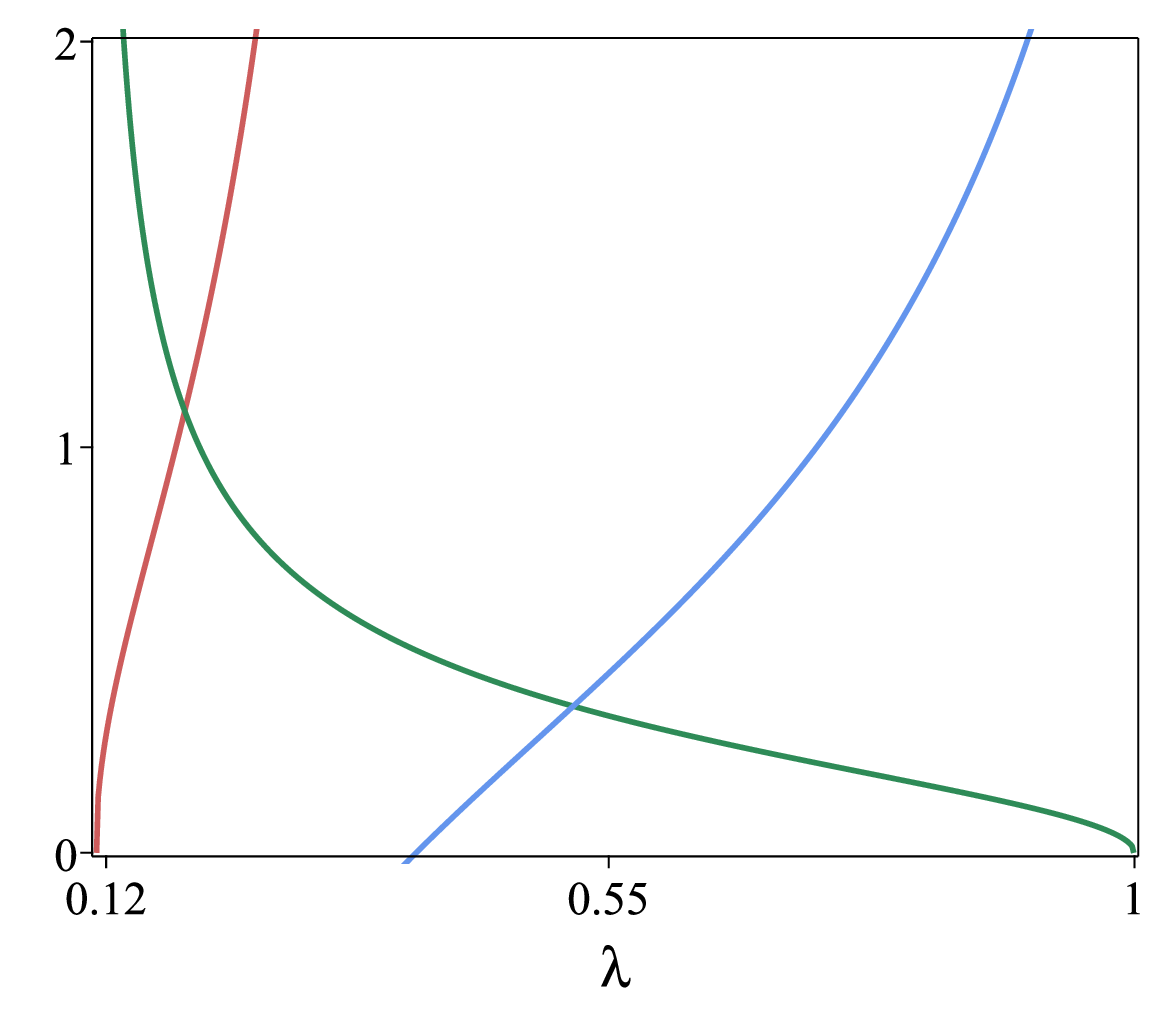}\includegraphics[width=0.33\linewidth]{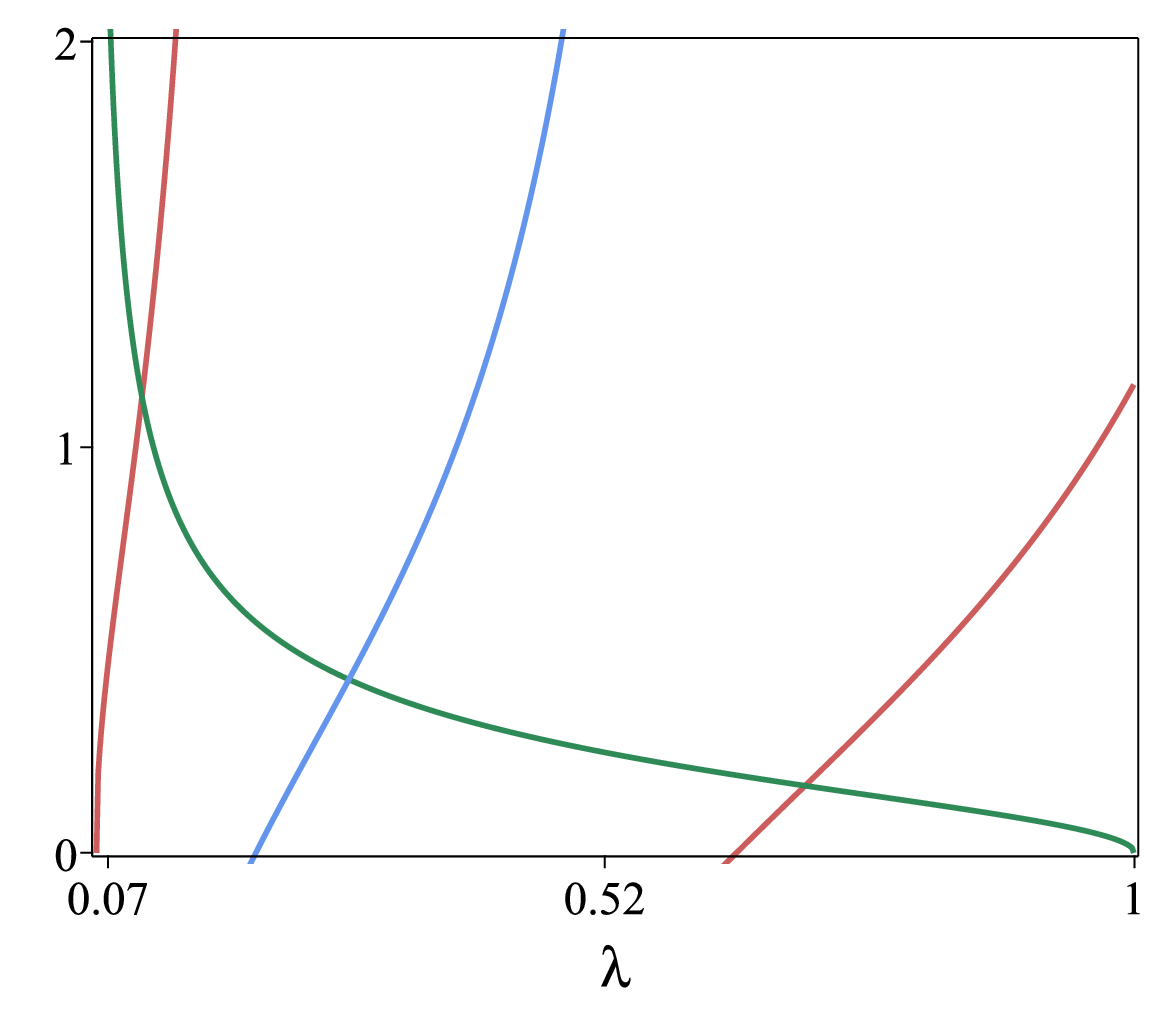}
    \includegraphics[width=0.33\linewidth]{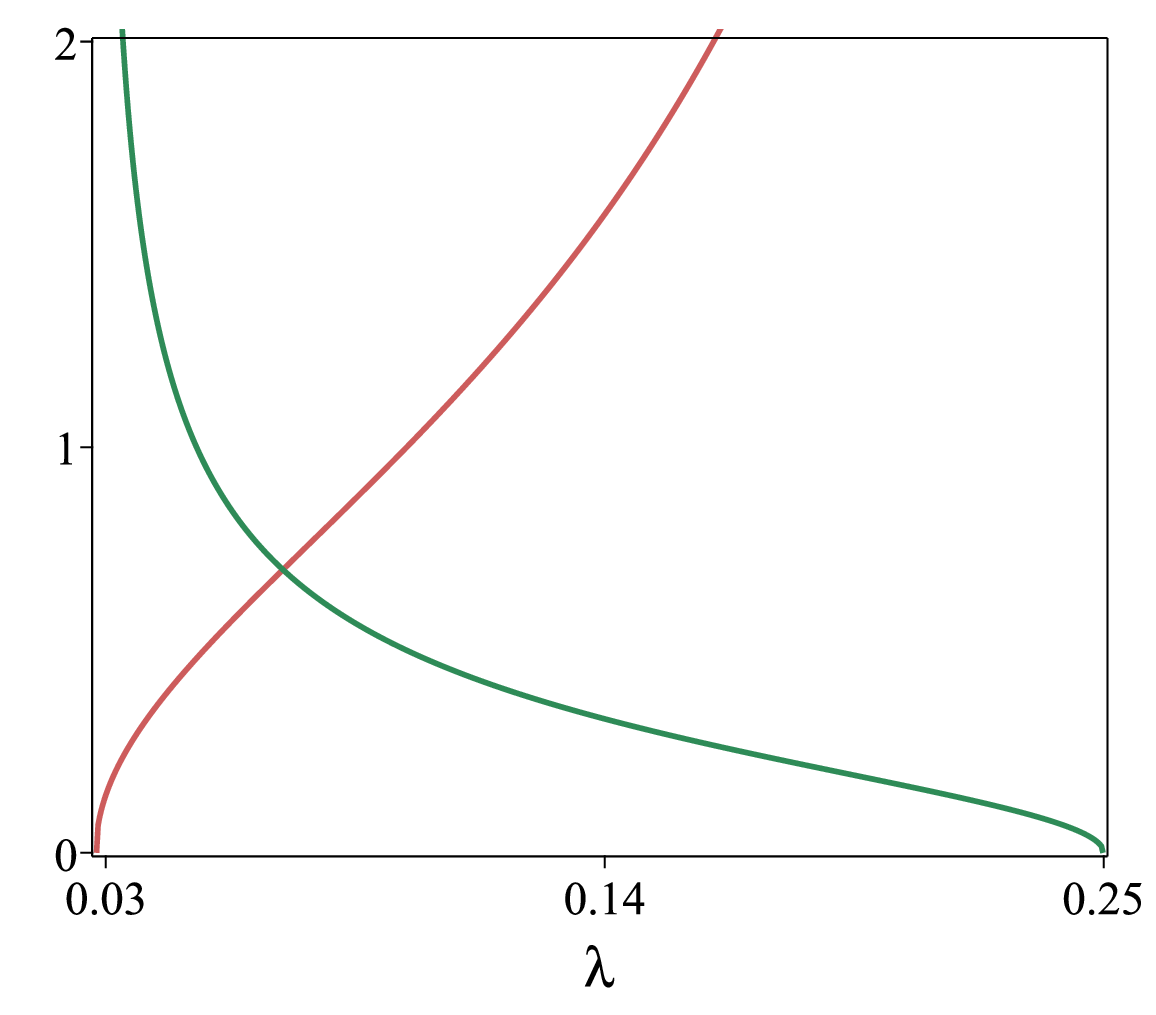}\includegraphics[width=0.33\linewidth]{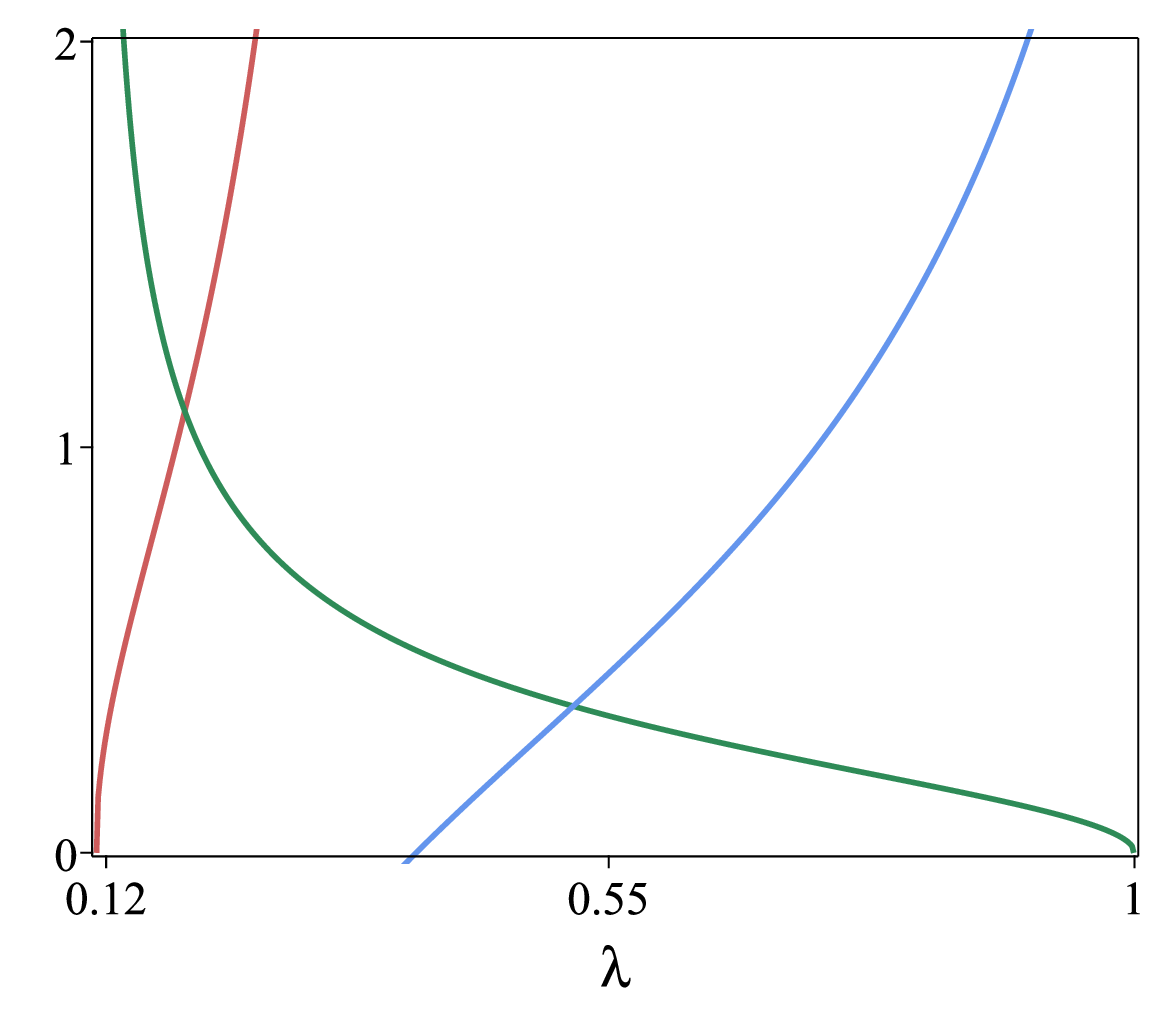}\includegraphics[width=0.33\linewidth]{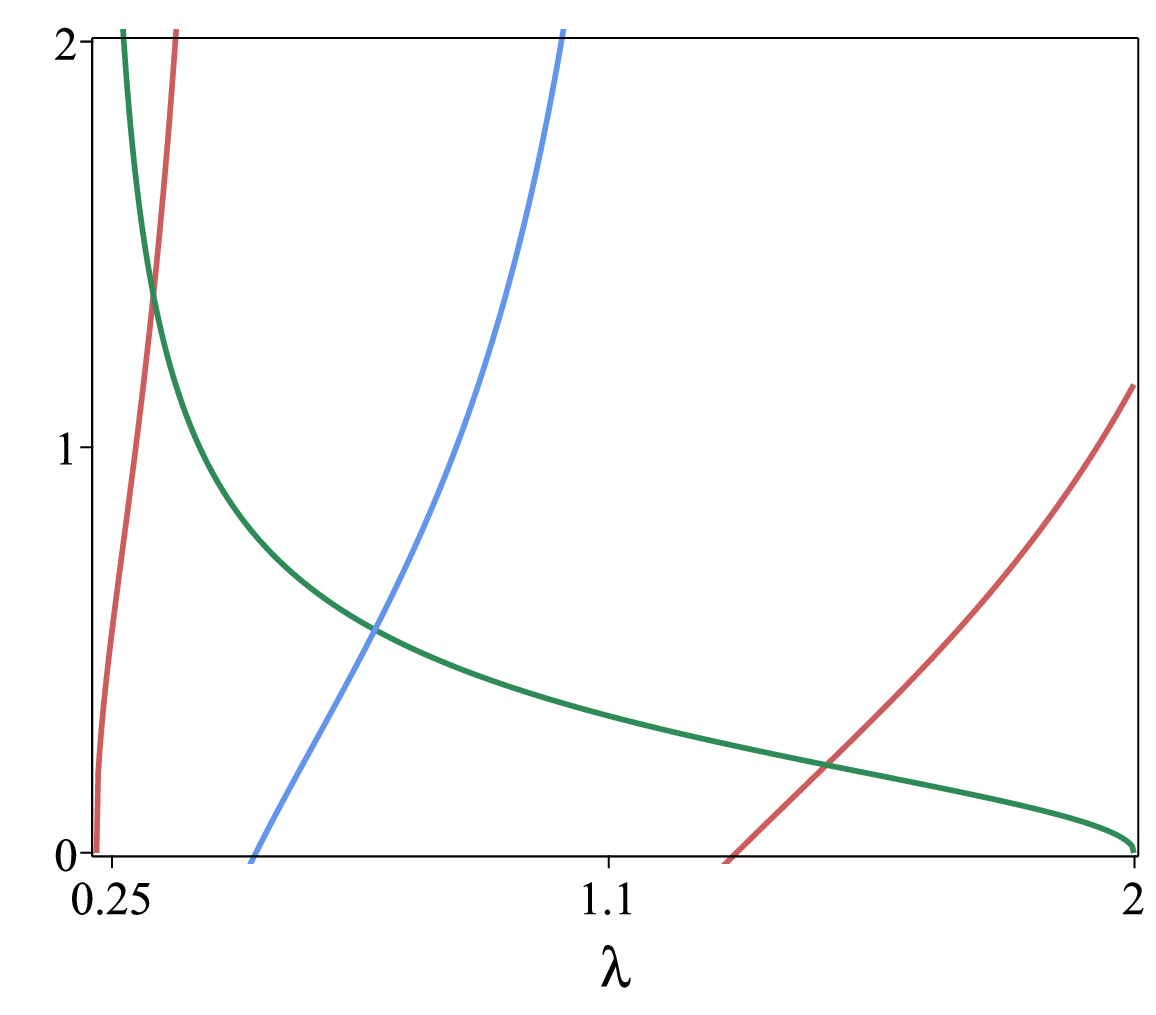}
    \caption{The functions $f_1$ (green), $f_2$ (red) and $f_3$ (blue) in Eqs.~\eqref{fs}. In the top row, we depict them for $\gamma=L=A=m=1$ and $\delta=1$ (top-left panel), $8$ (top-middle panel) and $16$ (top-right panel). In the bottom row, we display them for $\gamma=L=A=1$, $\delta=8$ and $m^2=1/4$ (bottom-left panel), $1$ (bottom-middle panel) and $2$ (bottom-right panel). For convenience, we have adopted $\lambda=\omega^2$ in the label of each horizontal axis. Notice that the range of $\lambda$ is not the same in each panel, as expected from the condition above Eqs.~\eqref{fs}.}
    \label{figf123}
\end{figure}

Before investigating the stability equation \eqref{stabchi} with the weight function \eqref{sigmasol}, we consider a simpler situation, described by
\be\label{toysigma}
\sigma(y) = \begin{cases}
    A+\delta, & |x|<L,\\
    A, & |x|>L,
\end{cases}
\ee
where $A$ is a positive parameter and $\delta$ is real, obeying $\delta>-A$. This function represents a toy model for a barrier if $\delta$ is positive and a hole if $\delta$ is negative. The eigenvalue equation \eqref{stabchi} associated to the above function only supports bound states for $\delta>0$, which is our current interest. More specifically, outside the interval in which $\delta$ makes the barrier, the eigenstates decay exponentially and, inside the barrier ($|x|<L$), they oscillate. We have found that the bound states arise for $m^2/(A+\delta)<\omega^2<m^2/A$, with the condition $f_1=f_2$ for the even states and $f_1=f_3$ for the odd ones, where
\bes\label{fs}
\bal
& f_1 = \sqrt{\frac{m^2-\omega^2A}{m^2-\omega^2(A+\delta)}},\\
& f_2= \tan\left(\sqrt{m^2-\omega^2(A+\delta)}\,L\right),\\
& f_3 = -\cot\left(\sqrt{m^2-\omega^2(A+\delta)}\,L\right).
\eal
\ees
In Fig.~\ref{figf123} we plot the above functions for several values of the parameters. The bound modes arise in the points where $f_2$ and $f_3$ intercept $f_1$. One can see that the number of bound states may increase with $\delta$ or $m^2$. Moreover, by investigating the behavior of these functions, one can show that the number of bound states may also increase with $L$.

The weight function \eqref{toysigma} allowed us to understand the basic features that may appear in the study of the eigenvalue equation \eqref{stabchi}, as it is described by analytical solutions. Let us now consider the weight function \eqref{sigmasol}, whose stability equation requires the use of special functions due to the presence of the hyperbolic secant in $\sigma(y)$. Surely, the nonlinear form of $\sigma(y)$ makes the investigation more intricate, but we may use the results of the function \eqref{toysigma} as a guide. We then turn our attention to the eigenvalue equation \eqref{stabchi} combined with \eqref{sigmasol}, which reads
\be\label{stabchisol}
-\xi_{yy} + m^2\xi(y)= \omega^2\left(\cfrac{2\sinh\gamma\,\sech^2(y)}{2Ce^{\gamma}-\sech^2(y)}+ e^{\gamma}\right)\xi(y).
\ee
We seek solutions that represent bound states. Even though the weight function makes this problem more complicated than the standard Schr\"odinger-like equation, it supports analytical solutions. For the even eigenfunctions, it is
\bes\label{xisols}
\be\label{xieven}
\begin{aligned}
\xi_{even}(y) &= \sech^{u}(y)\,\text{HeunG}\bigg(v,\frac{u^2+u}4+\frac{v}2\omega^2\sinh\gamma,\\
&u,\frac{1+u}{2},\frac12,0,v\tanh^2(y)\bigg),
\end{aligned}
\ee
and, for the odd ones, we get
\be\label{xiodd}
\begin{aligned}
\xi_{odd}(y) &= \sech^{u}(y)\tanh(y)\times\\
&\text{HeunG}\bigg(v,\frac{(u+1)(u+2)}4+\frac{v}2\omega^2\sinh\gamma,\\
&\frac{1+u}2,1+\frac{u}{2},\frac32,0,v\tanh^2(y)\bigg).
\end{aligned}
\ee
\ees
where $u=\sqrt{m^2-e^\gamma\omega^2}$, $v=1/(1-2Ce^{\gamma})$ and $\text{HeunG}$ denotes the Heun general function. Of course, only some values of of $\omega$ are compatible with the boundary conditions $\xi(\pm\infty)=\xi_y(\pm\infty)=0$, which are the ones of our interest.

\begin{figure}[t!]
    \centering
\includegraphics[width=0.25\linewidth]{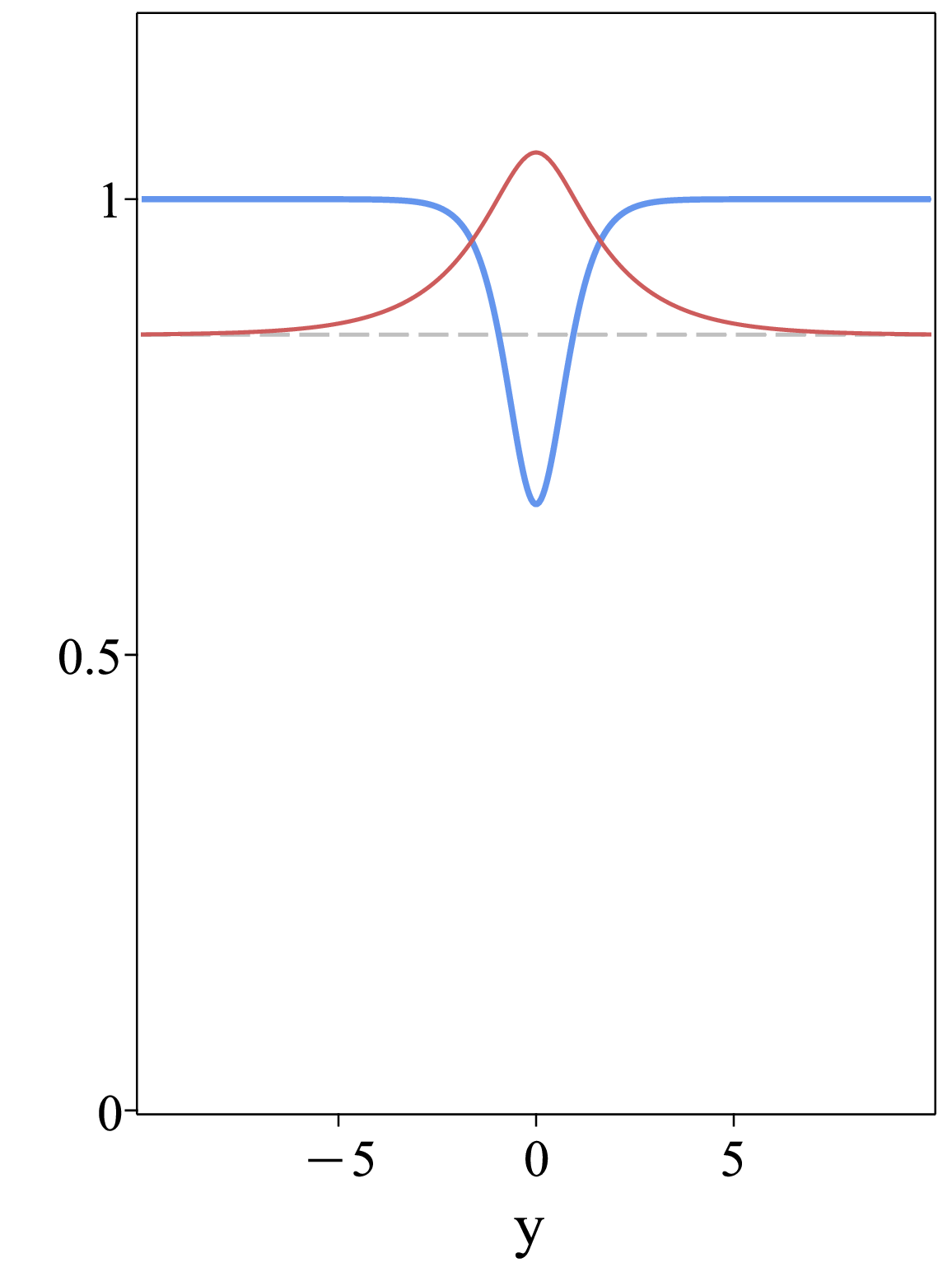}\includegraphics[width=0.25\linewidth]{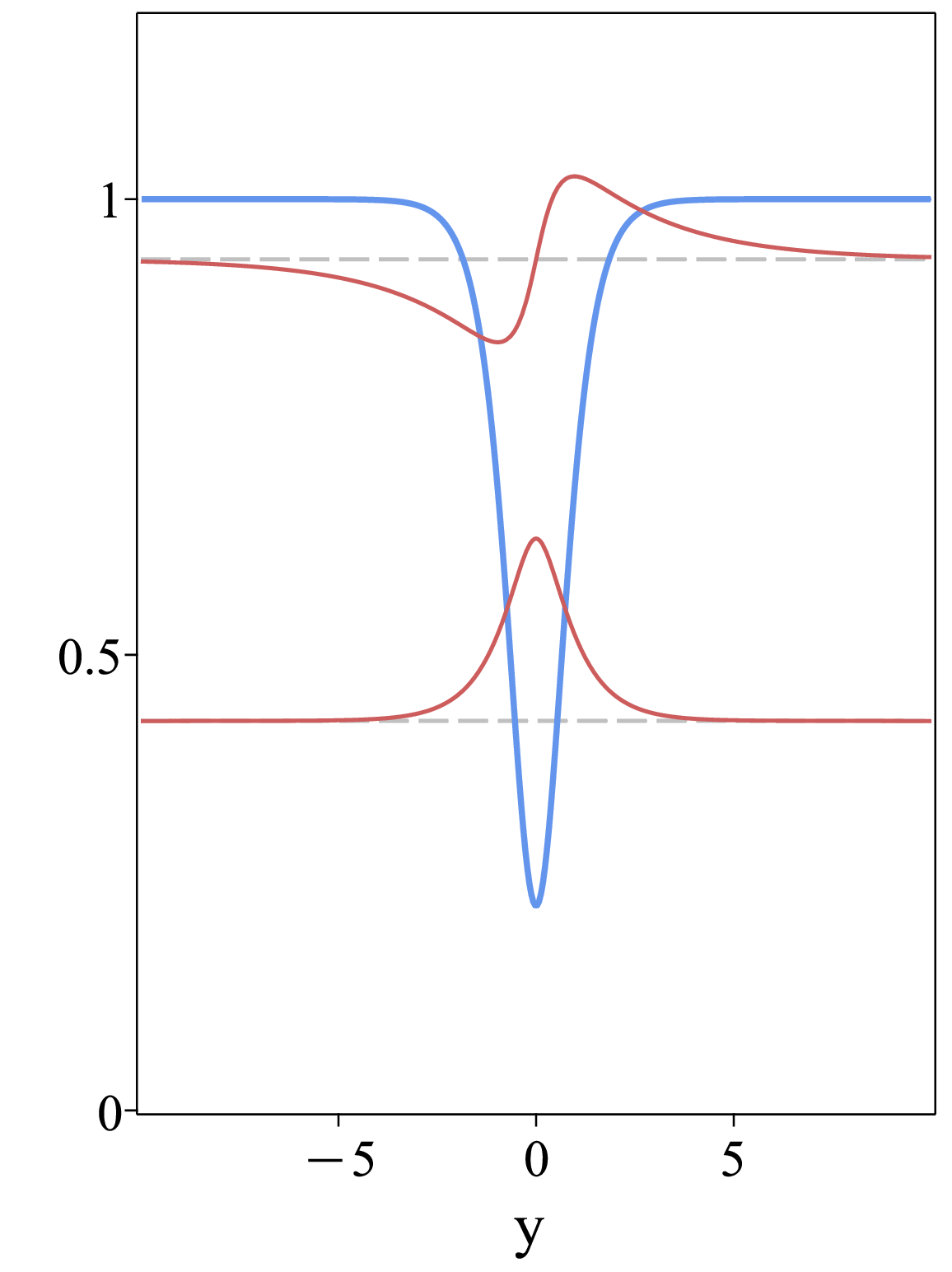}\includegraphics[width=0.25\linewidth]{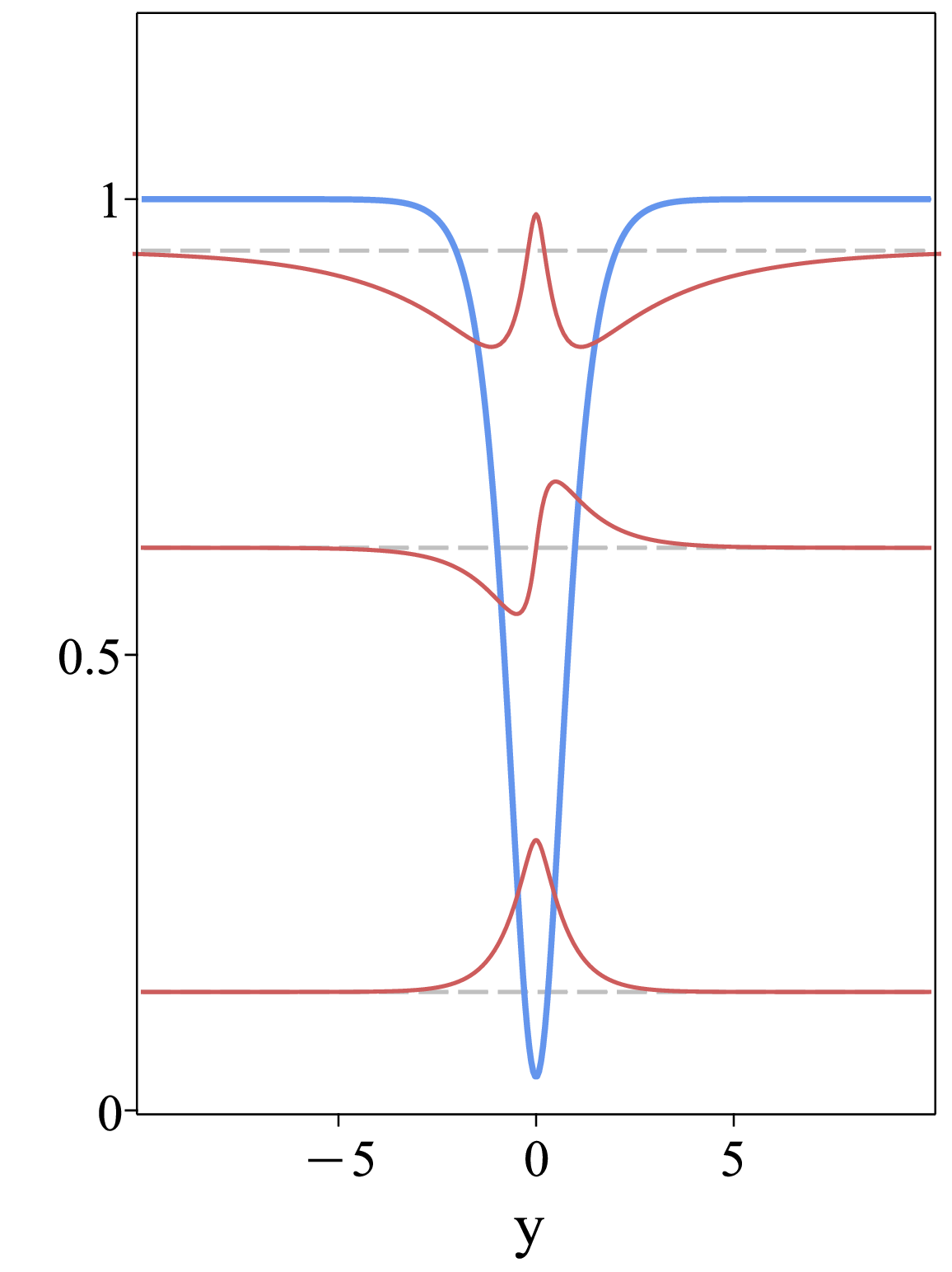}\includegraphics[width=0.25\linewidth]{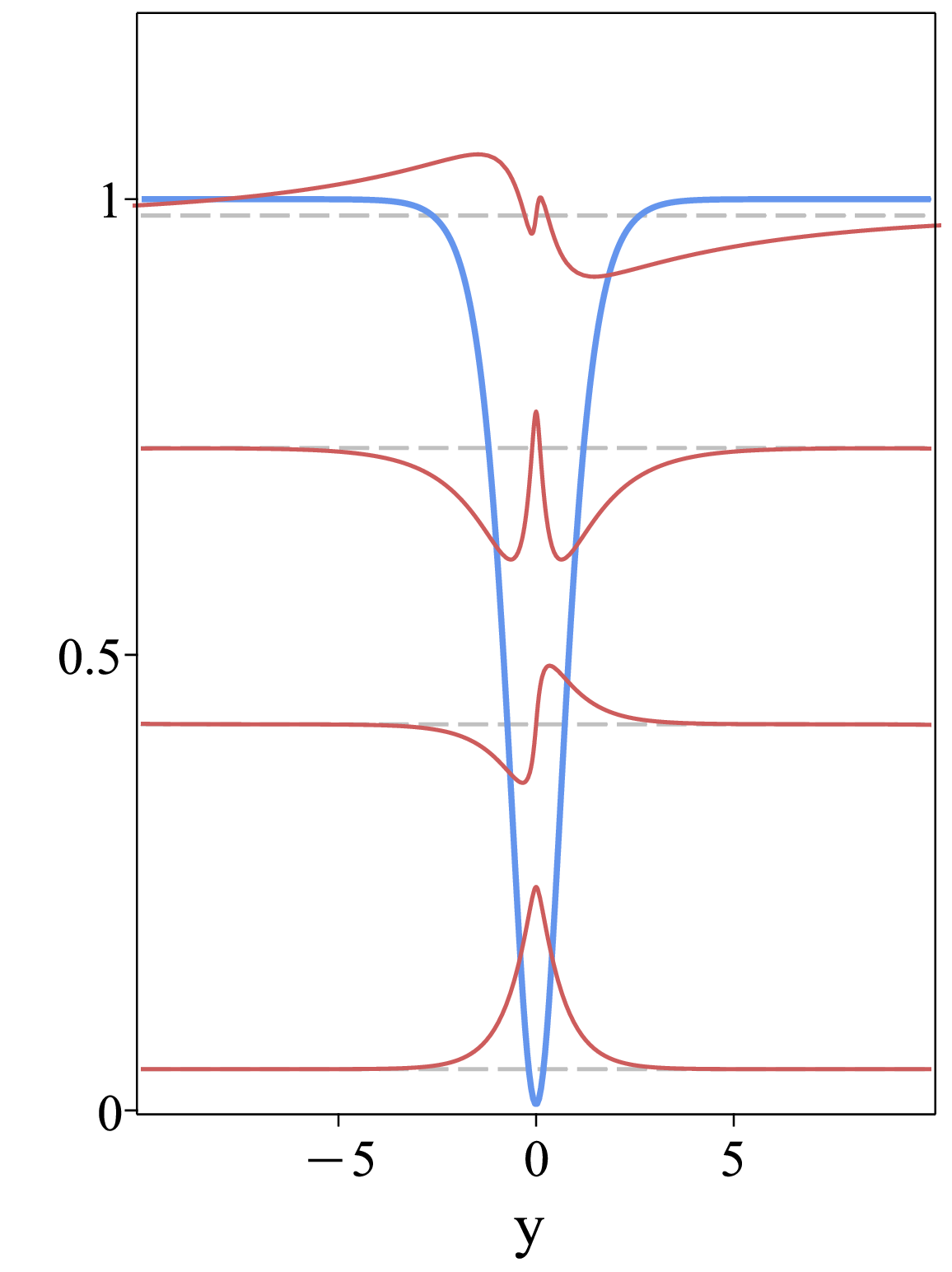}
    \includegraphics[width=0.25\linewidth]{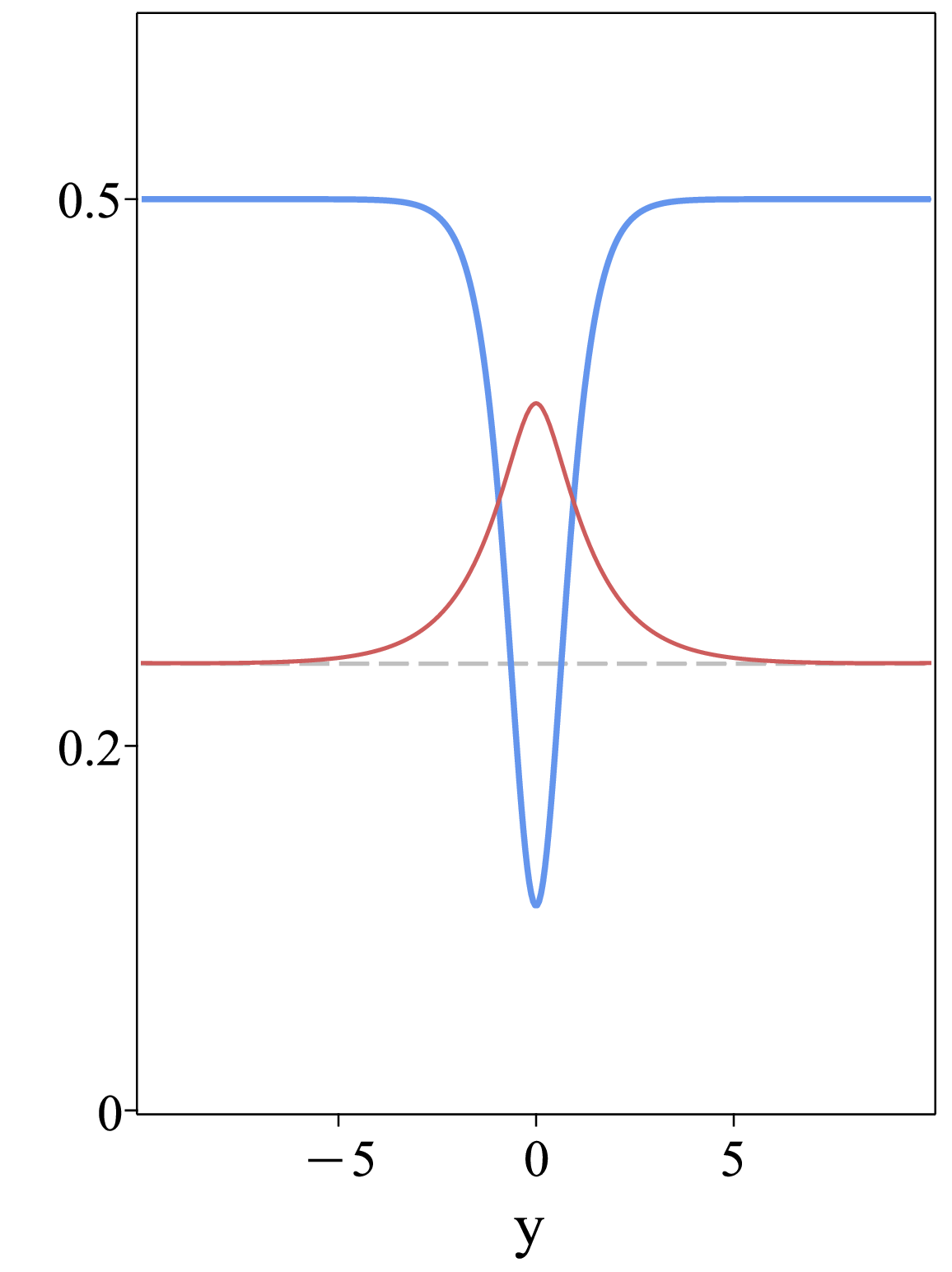}\includegraphics[width=0.25\linewidth]{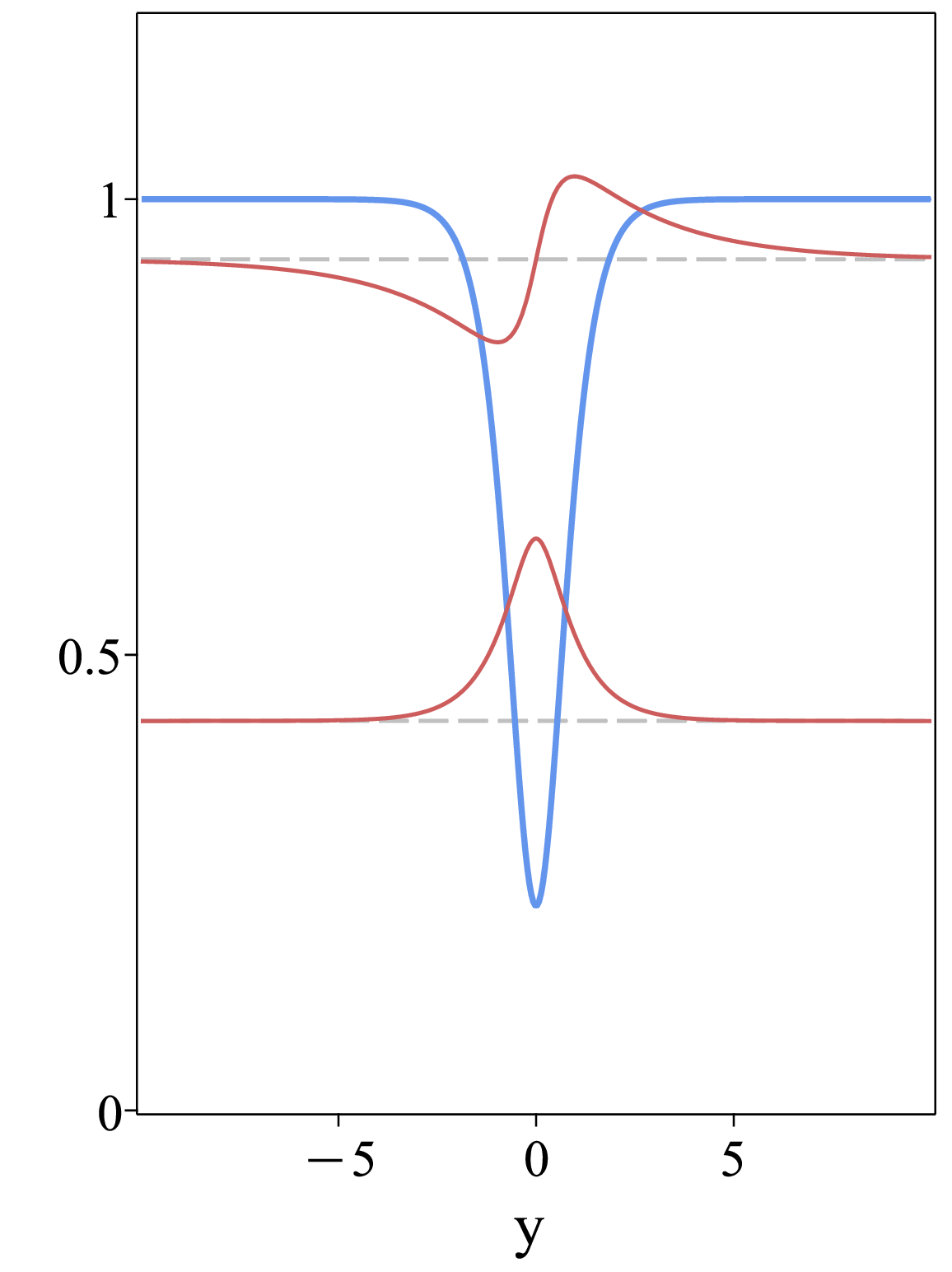}\includegraphics[width=0.25\linewidth]{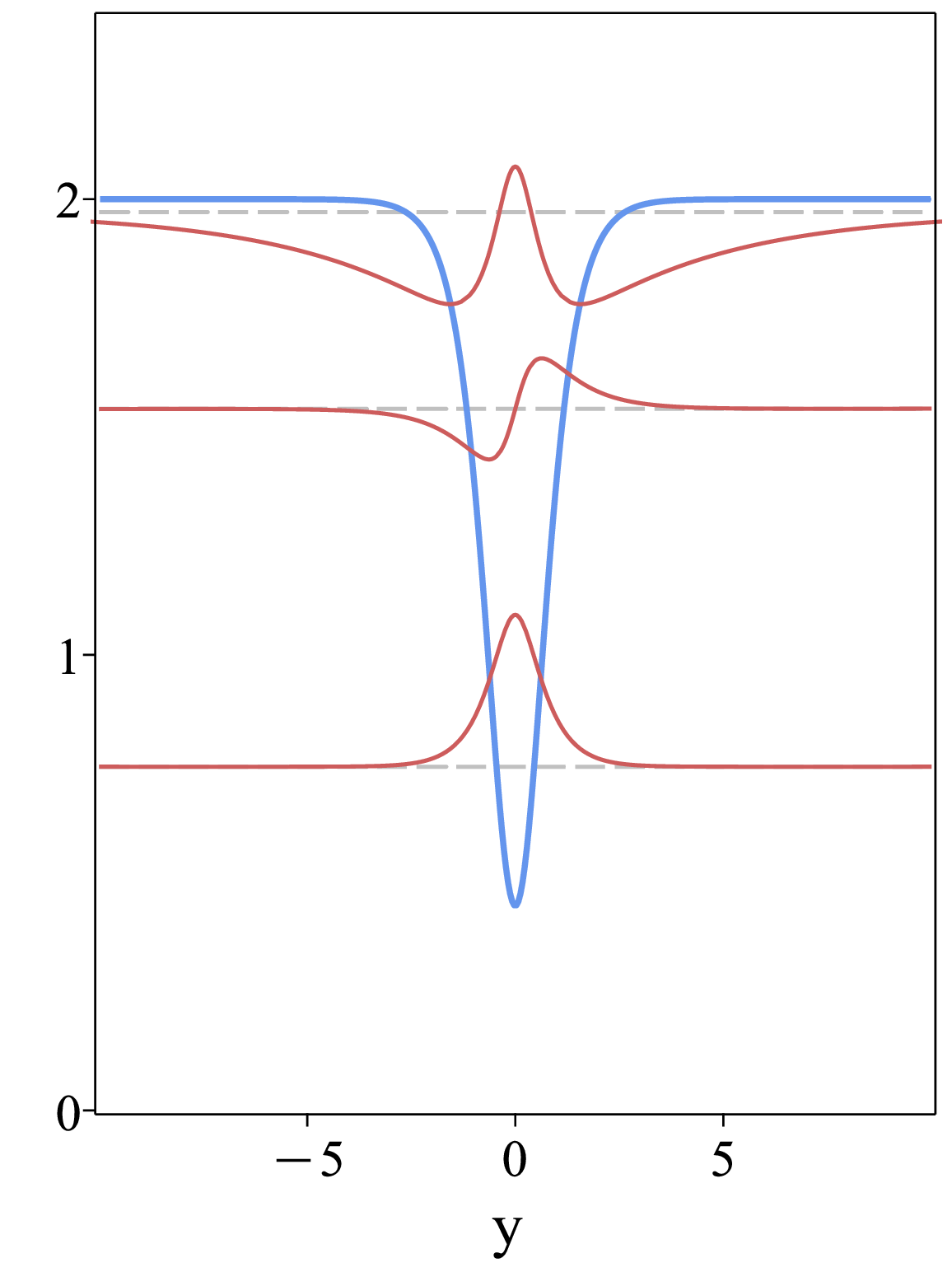}\includegraphics[width=0.25\linewidth]{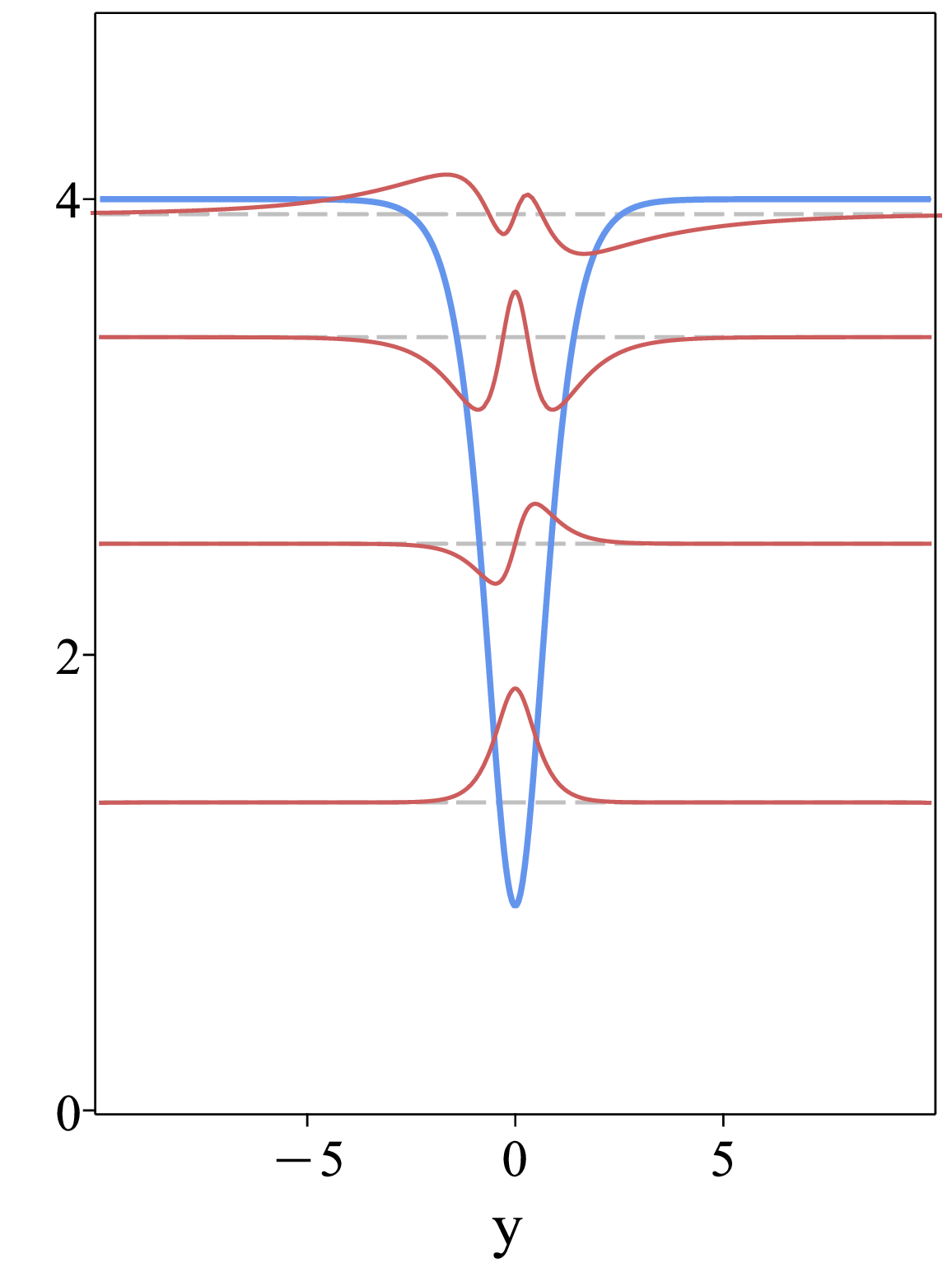}
    \caption{The ratio $m^2/\sigma(y)$ (blue, solid lines) and the eigenfunctions (red, solid lines) associated to the eigenvalues (gray, dashed lines). In the top row, we consider $\gamma=1$, $m^2=e$ and $C=0.5$ (top left), $0.23$ (top middle-left), $0.19$ (top middle-right) and $0.185$ (top right). In the bottom row, we consider $\gamma=1$, $C=0.23$ and $m^2=e/2$ (bottom left), $e$ (bottom middle-left), $2e$ (bottom middle-right) and $4e$ (bottom right). The eigenfunctions obey Eqs.~\eqref{xisols}. For better visualization, the vertical axes are not the same in the bottom panels.}
    \label{figmodos}
\end{figure}

To understand the behavior of the states, we display the bound states from above eigenfunctions and their respective eigenvalues in Fig.~\ref{figmodos} for some values of the parameters. We can see that only specific discrete values of $\omega$ lead to bound states. Notice that, in order to ensure that the above eigenfunctions are real, the eigenvalues must obey $\omega^2<m^2e^{-\gamma}=m^2/\sigma_\infty$, as expected from the general discussion in the paragraph right above Eq.~\eqref{sigmasol}. However, the asymptotic behavior of $\sigma(y)$ is not enough to ensure the presence of bound states. As we have also commented in the aforementioned paragraph, we have to take into account another condition, related to the points $y_*$ in which $\omega^2> m^2/\sigma(y_*)$. By investigating this expression, we have found that, in the core of the kink \eqref{sol}, these points may exist if
\be
\frac{2Ce^\gamma-1}{2Ce^{2\gamma}-e^{-\gamma}}<\frac{\omega^2}{m^2}<e^{-\gamma},
\ee
in which the inequality in the right came from the condition that emerged from the asymptotic behavior of $\sigma(y)$. Indeed, all the eigenfunctions depicted in Fig.~\ref{figmodos} have eigenvalues obeying the above expression. 

We then analyze the eigenfunctions and eigenvalues displayed in Fig.~\ref{figmodos}. In the top row, we show how the system behaves when we change the parameter $C$ while $\gamma$ and $m$ are fixed. We emphasize that, since we are considering $\gamma=1$ and $m^2=e$, we must take $C>1/(2e)\approx0.18394$ to ensure that the weight function \eqref{sigmasol} is always positive. Therefore, we start with $C=0.5$, for which there is only a single bound state. As we decrease $C$, the number of bound states may increase; we have found four of them for $C=0.185$. In the bottom panels of the figure, we vary the parameter $m$ and fix the others. In this situation, we see that $m^2=e/2$ leads to a single bound state. As $m$ gets larger, one may obtain more bound states. Indeed, for $m^2=4e$, we get four of them. It is also worth to remark that, as expected, the $k-$th bound state engenders $k-1$ nodes, so the ground state is always free of nodes. Since negative eigenvalues are absent, the vacuum solution $\chi=0$ is stable.

We highlight that the presence of the bound modes in the fluctuations around the vacuum solution $\chi=0$ is due exclusively to the scalar ModMax model \eqref{LL}, whose signature in the stability equation appears in the weight function $\sigma(y)$ that couples the fields nontrivially.

In this Letter, we have studied a dimensionally-reduced version of the ModMax model, called scalar ModMax, as it depends on the scalar fields $\phi$ and $\chi$. We have considered the presence of a constant electric field in the system and focused on the case in which one of the scalar fields is in the vacuum, $\chi=0$. In the static case, we show that the equation of motion that governs the non-constant field can be mapped into the canonical model by a redefinition of the spatial coordinate that depends on the parameter of the ModMax model, $\gamma$. By perturbing the static fields, we show that the each eigenvalue equation are decoupled in the fluctuations and the perturbation around the solution $\chi=0$ depends explicitly on $\phi$ via the weight function $\sigma(y)$ that arises in the Sturm-Liouville eigenvalue equation. It is worth commenting that the study of small fluctuations around field configurations with constant electric fields or constant scalar gradients has previously been used in several models with non-analytic kinetic terms; see Refs.~\cite{small1,small2}. To better understand the features associated to the number of bound states of this equation, we have first investigated a toy model in which $\sigma(y)$ is a rectangular barrier. This allowed for an analytical treatment from which we could take advantage to show that bound states may arise in specific values of the parameters in a range of the eigenvalues. Next, we have studied the case in which the weight function is described by the sine-Gordon solutions in $\phi(y)$. This situation is more intricate and requires the use of Heun functions. Even so, the fluctuations are described by analytical functions and we have found that bound states arise for specific values of the parameters in a $\gamma$-dependent range of $\omega^2$. The system is linearly stable, as the ground state engenders no nodes.

As perspectives, one may consider the situation in which the potential $U(\phi,\chi)$ couples the fields and/or the case with both fields being non uniform, supporting topological and non-topological structures. Another line of continuation of the current paper follows from the expansion of the scalar ModMax Lagrangian density \eqref{LL} for very large values of $C$. In this regime, one gets $\LL \approx \cosh\gamma\,X + \sinh\gamma\,\left(C+X+Y^2/2C\right)$, such that the dynamical terms $X$ and $Y$ do not mix, with $X$ and $Y$ being linear and quadratic, respectively. In this direction one may consider, in future studies, Lagrangian densities in which terms in $Y$ are generalized, in the form $\LL=X + F(Y)$, or dimensionally-reduced versions of the models studied in \cite{modmax6}, to investigate how they impact the study of kinks and their linear stability. Another perspective concerns the model \eqref{LL} with null electric field ($C=0$), which requires a careful approach, since the vacuum of the system becomes ill defined.

One may also study the case in which the solution $\phi(y)$ has a different profile from the one in Eq.~\eqref{sol}, supporting internal structure or tails with distinct behavior, impacting the weight function \eqref{sigma} and modifying the study of the bound states. In this case, one may also transform the Sturm-Liouville equation into a Schr\"odinger-like one; the use of numerical methods may also be required in the process.  Some of these issues are currently being considered and will be reported elsewhere.

\acknowledgments{We acknowledge financial support from the Brazilian agencies Conselho Nacional de Desenvolvimento Cient\'ifico e Tecnol\'ogico (CNPq), grants Nos. 309092/2022-1 (FAB), 402830/2023-7 (MAM and RM), 306151/2022-7 (MAM), 310994/2021-7 (RM) and 304290/2020-3 (EP).}



\end{document}